\documentstyle[12pt]{article}
\renewcommand{\big}{}
\begin{document}
\begin{center}
\bf{Exact ground and excited states of a t-J ladder doped with two holes }
\vskip 1cm

                     Indrani Bose and Saurabh Gayen*\\
                    Department of Physics\\
                    Bose Institute\\
                    93/1, A.P.C.Road\\
                    Calcutta-700 009, India.\\
                   $\ast$ Physics Department\\
                    Scottish Church College\\
                    Calcutta-700006
\end{center}
\begin{abstract}
A two chain ladder model is considered described by the strong coupling 
$t-t^\prime-J-J^\prime$ Hamiltonian.  For the case of two holes moving in
 a background of antiferromagnetically interacting spins, exact,
analytical results are derived for the ground state energy and low-lying 
excitation spectrum. The ground state is a bound state of two holes with 
total spin S=0. The charge excitation is gapless and the spin excitation 
has a gap. The corresponding wavefunctions are also exactly determined. The 
bound hole pair is found to have symmetry of the d-wave type. In the limit of  
strong rung coupling, the model maps onto an effective hard core boson model 
which exhibits dominant superconducting pairing correlations.
\end{abstract}

\section*{I. Introduction}
In the last few years, ladder systems have been studied 
extensively \cite{Dagotto1,Rice,Bose1}. 
Interacting electron systems in one dimension (1d) are more or less well 
understood. There are several rigorous results available for such systems 
\cite{Bose1}. Powerful techniques like the Bethe Ansatz \cite{Bethe} and 
bosonization \cite{Jv} have yielded much useful information about such 
systems. After the discovery of high-$T_{c}$ cuprate superconductivity, 
2d interacting electron systems acquired new significance due to the fact 
that the dominant electronic and magnetic properties of the cuprate systems 
are associated with the $CuO_2$ plane \cite{Dagotto2,AP}. There are, however, 
very few rigorous results available for 2d systems. Ladders, consisting of 
n-chains coupled by rungs, interpolate between 1d and 2d and their study is 
expected to be useful for a proper understanding  of interacting many body 
systems. The possibility of deriving rigorous results is also more. 
A number of ladder systems have been discovered recently exhibiting a variety
of interesting phenomena\cite{Dagotto1,Rice,Bose1}. Physical insight obtained from the study of ladders
is also expected to be relevant for high-$T_c$ cuprate systems. The cuprates,
in the spin-disordered phase, are doped spin liquids. Below optimal doping
levels and well above the superconducting transition temperature $T_c$, there
are experimental signatures of a spin gap(SG)\cite{AP} opening up. The 'gap' has been
ascribed to pre-formed Cooper pairs of holes which lack the long-range phase
coherence of the superconducting state. The Cooper pairs become phase coherent
only below $T_c$ giving rise to superconductivity. Dagotto et al\cite{Dagotto3} were the first
to show that a two-chain ladder has a spin liquid ground state and a SG in the
excitation spectrum. On doping the system with holes, binding of holes in pairs
is possible, giving rise to dominant superconducting(SC) pairing correlations.
A few years later, a hole doped two-chain ladder system $Sr_{0.4}Ca_{13.6}Cu_{24}O_{41.84}$
 was discovered which exhibits superconductivity under pressure\cite{Maekawa}.

 The relationship between `pseudo'  spin-gap, pre-formed hole pairs and
 superconductivity is not well-understood in the case of cuprate systems. For
 the ladder system, the SG is a real gap and the binding of holes leading to
 SC pairing correlations can be explicitly demonstrated. Resistivity
 measurements of the ladder compound $(Sr,Ca)_{14}Cu_{24}O_{41}$ show unusual
 temperature dependence as in the case of cuprates \cite{FF}highlighting further
 similarities between the two systems. Bose and Gayen \cite{Bose2,Bose3,Bose4,Bose5}have constructed a
 two-chain t-J type ladder model for which several exact, analytical results
 can be derived in the undoped as well as doped cases. For two holes, the
 possibility of binding of holes was suggested but the bound state
 spectrum was not derived. In Section II of this paper, we give a detailed
 derivation of the low-lying spin and charge excitation spectrum of the ladder model
 in the two-hole sector. We show that the ground state consists of a bound
 pair of holes. The spin excitation spectrum has a gap and the charge excitation
 is gapless. The two-hole wave functions are also computed. The two-hole
 bound state wave function is shown to have modified d-wave symmetry. All
 these results are exact and analytic in nature. The dominance of SC pairing
 correlations in the ladder model is shown in an approximate, analytical
 manner.

\section*{II. Exact two-hole excitation spectrum}
The two-chain ladder model consists of two chains, 
each described by a t-J Hamiltonian, coupled by $t^\prime-J^\prime$ interactions 
between them (Fig.1). The model is described by the $t-t^\prime-J-J^\prime$ Hamiltonian:  
\begin{eqnarray}
H\,&=&-\,\sum_{i,j,\sigma}t_{ij}\,(1-\,n_{i-\sigma}\,) C_{i,\sigma}^{\dag}\,C_{j,\sigma}\,(1-\,n_{j-\sigma}) + H.C.+\,\sum_{<ij>}J_{ij}\,\vec{S_i}\cdot\vec{S_j}\nonumber\\ 
   &=&\,H_{t} + H_{t^\prime} + H_{J} + H_{J^\prime} 
\end{eqnarray}                                                         
The constraint that no site can be doubly occupied is implied in the model.
The hopping integral $t_{ij}$ has the value t for nearest-neighbour(NN) hopping 
within a chain and also for diagonal transfer between chains (solid lines in Fig.1).
The corresponding spin-spin interactions $J_{ij}$ are of strength J. The spins 
have magnitude 1/2. The hopping integral across vertical links (broken lines)
connecting two chains has the strength $t^\prime$. The corresponding spin-spin interaction 
strength $J_{ij}$ is $J^\prime$. We assume t and $t^\prime$ to be positive. In the conventional 
two-chain spin-ladder, the diagonal interaction and hopping terms are absent.
The inclusion of the diagonal terms of the same strength
as the intra-chain ones enables one to reduce the difficult N-body problem to an
easily solvable few body problem. The conventional spin ladder model, in the absence 
of diagonal terms, constitutes a many-body problem for which no simplification 
occurs. The only exact results, which are available, are numerical results based  
on exact diagonalization of finite ladders \cite{Dagotto1,Rice,Troyer}.  

In the half-filled limit, i.e, in the absence of holes, the $t-t^\prime-J-J^\prime$-Hamiltonian 
in (1) reduces to $H_{J}$+$H_{J^\prime}$. The exact ground state $\psi_g$ ( for $J^\prime\ge2J$ )
consists of singlets along the rungs of the ladder \cite{Bose2}. The ground state
energy $E_g=-(3J^\prime/4)N$, where N is the number of rungs. An exact excited state 
can be constructed by replacing a singlet by a triplet. Creation of a triplet 
costs an amount of energy $J^\prime/4$ so that the spin gap $\Delta_{SG}=J^\prime$. The 
excitation is localised and has no dynamics. Let us now consider the case of
a single hole doped into the ladder. In the presence of holes a single rung 
can exist in nine possible states: (i) empty, (ii) two bonding states, (iii) two anti-bonding 
states, (iv) one singlet state and (v) three triplet states. These states are shown 
below.
\begin{eqnarray*}
&&(i)\,\,\left({{}^{\big O}_{\big O}}\right)\,\,\,\,\,\,\,\,(ii)\,\,\frac{1}{\sqrt{2}}\left({{}^{\big \uparrow}_{\big O}}\,+\,{{}^{\big O}_{\big \uparrow}}\right)\,\,\,\,,
\,\,\,\frac{1}{\sqrt{2}}\left({{}^{\big \downarrow}_{\big O}}\,+\,{{}^{\big O}_{\big \downarrow}}\right)\\
&&(iii)\,\,\,\,\frac{1}{\sqrt{2}}\left({{}^{\big \downarrow}_{\big O}}\,-{{}^{\big O}_{\big\downarrow}}\right)\,,\,\,\,\,\,\,\frac{1}{\sqrt{2}}\left({{}^{\big\uparrow}_{\big O}}\,-\,{{}^{\big O}_{\big\uparrow}} \right)\\
&&(iv)\frac{1}{\sqrt{2}}\left({{}^{\big\uparrow}_{\big\downarrow}}\,-\,{{}^{\big\downarrow}_{\big\uparrow}}\right)\,\,\,\,\,(v)\,\,{{}^{\big\uparrow}_{\big\uparrow}}\,\,,\,\,{{}^{\big\downarrow}_{\big\downarrow}}\,\,,\,\,
\frac{1}{\sqrt{2}}\left({{}^{\big\uparrow}_{\big\downarrow}}\,+\,{{}^{\big\downarrow}_{\big\uparrow}}\right)
\end{eqnarray*}

 A single hole hops in a background of antiferromagnetically 
 interacting spins. This, in general, is a difficult many body problem because
 as the hole hops it gives rise to spin excitations in the system. The
 inclusion of diagonal hopping terms in our model leads to a cancellation of
 all the terms containing spin excitations, resulting in a perfect, coherent
 motion of the hole. We illustrate this through an explicit example. 
 Consider a single hole in a bonding state, located in the m-th rung. All
 other rungs are in singlet spin configurations. A pictorial representation
 of the state is
\begin{equation}
\big|\,\,\big|\,\,\big|\cdot\cdot\cdot\frac{1}{\sqrt{2}}\left({}^\uparrow_O+{}^O_\uparrow\right)_{m}\big|_{{}_{(m+1)}}
\cdot\cdot\cdot
\end{equation}                   
\begin{equation}
\frac{1}{\sqrt{2}}\left({}^\uparrow_O+{}^O_\uparrow\right)_{m}\big|_{{}_{(m+1)}}\,\equiv\,
\frac{1}{2}\left({}^{\uparrow\,\uparrow}_{O\downarrow}-{}^{\uparrow\,\downarrow}_{O\uparrow}
+{}^{O\uparrow}_{\uparrow\,\downarrow}-{}^{O\downarrow}_{\uparrow\,\uparrow}\right)
\end{equation}
 The state is an exact eigenstate of the $J,J^\prime,t^\prime$ part of the $t-t^\prime-J-J^\prime$ Hamiltonian. 
 Let us now apply $H_{t}$ on the state. Since background electrons are fermions, 
 their ordering is important and one has to keep track of signs during 
 interchanges. 
The ordering of fermions follows the convention
\begin{center}
         \,1\,\,\,\,\,\,\,3\,\,\,\,\,\,\,5\,\,\,\,\,\,$\cdot$\,\,\,\,\,\,$\cdot$\\
          2\,\,\,\,\,\,\,4\,\,\,\,\,\,\,6\,\,\,\,\,\,\,$\cdot$\,\,\,\,\,\,\,$\cdot$\\
\end{center}
  On operating with $H_{t}$ on the state shown in (3), one gets
\begin{eqnarray*}
 H_t\,\left({}^{\uparrow\,\uparrow}_{O\,\downarrow}\right)\,&=&\,\,\,\,\,t\,{}^{\uparrow\,\uparrow}_{\downarrow\,O}\,-\,t\,{}^{\uparrow\,O}_{\uparrow\,\downarrow}\\
-H_t\,\left({}^{\uparrow\,\downarrow}_{O\,\uparrow}\right)\,&=&\,-t\,{}^{\uparrow\,\downarrow}_{\uparrow\,O}\,+\,t\,{}^{\uparrow\,O}_{\downarrow\,\uparrow}\\
* H_t\,\left({}^{O\,\uparrow}_{\uparrow\,\downarrow}\right)\,&=&\,\,\,\,\,t\,{}^{\uparrow\,O}_{\uparrow\,\downarrow}\,-\,t\,{}^{\downarrow\,\uparrow}_{\uparrow\,O}\\
-H_t\,\left({}^{O\,\downarrow}_{\uparrow\,\uparrow}\right)\,&=&\,-t\,{}^{\downarrow\,O}_{\uparrow\,\uparrow}\,+\,t\,{}^{\uparrow\,\downarrow}_{\uparrow\,O}
\end{eqnarray*}

The states in the second column are obtained due to diagonal hopping
of the hole. There is a cancellation of the terms containing parallel spin
pairs and the final state is given by

\begin{equation}
{\big|}_{m}\,{\frac{1}{\sqrt{2}}}\,{\left({}^{\uparrow}_{O}+{}^{O}_{\uparrow}\right)}_{m+1}
\end{equation}

One finds that the hole accompanied by a free spin-1/2 moves coherently by one
lattice unit(compare with Eqn.(3)). The eigenvalue problem now becomes very easy
to solve. Let
\begin{equation}
\Psi(m)=\big|\,\big|\,\big|\cdot\cdot\cdot\frac{1}{\sqrt{2}}\left({}^\uparrow_O+{}^O_\uparrow\right)_{m}\big|_{{}_{(m+1)}}
\cdot\cdot\cdot
\end{equation}                   
\begin{equation}
\Psi\,=\,\frac{1}{\sqrt{N}}\,\sum _{m=1}^{N}\,e^{i\,k\,m}\,\Psi(m)
\end{equation}
$\Psi$ is an exact eigenstate of the $t-t^\prime-J-J^\prime$ Hamiltonian with the energy eigenvalue
\begin{equation}
\,E_1\,=\,2\,t\,cos(k)\,-\,t^\prime\,+\,3\,J^\prime/4
\end{equation}
The energy is measured with respect to that of the ground state energy in the 
undoped state. Refs.\cite{Bose2,Bose3} give a detailed discussion of the single hole spectrum
for both bonding and antibonding hole states. For conventional spin ladders, Troyer
et al \cite{Troyer} have found numerical evidence of quasi-particle (QP) excitations
carrying charge +e and spin-1/2. The charge and spin may be located on different  
rungs. In the exact eigenstate of Eqn.(6), the positively charged hole and the spin-1/2 are always
located on the same rung. We refer to the composite object as hole -QP.

Let us now consider the case of two holes. The two holes can be introduced on
the same rung or on separate rungs. Other rungs are in the singlet spin
configurations. If the holes are located on two separate rungs, there are two free spins which can 
combine to give either a triplet or a singlet. In the triplet sector, the two 
hole-QPs can scatter against each other giving rise to a continuum of scattering  
states with energy
\begin{equation}
\,E_{cont}\,=\,4\,t\,cos\,(K/2)\,cos\,q\,-\,2\,t^\prime\,+\,3\,J^\prime/2
\end{equation}

$K (=k_1+k_2)$ and $q (=(k_1-k_2)/2)$ are the centre of mass momentum and the 
relative momentum wave vectors. The two-hole ground state belongs to the singlet
sector. The exact eigenvalue equations have already been derived in Ref.[13] but a full
analysis of these equations has so far not been done. Define the wave functions
\begin{eqnarray}
\phi(m_1,m_2)&=&\frac{1}{2\sqrt{2}}\big[\big|\cdot\cdot\cdot
\big|\left({}^\uparrow_O+{}^O_\uparrow\right)_{m_{1}}\big|\cdot\cdot\cdot
\left({}^\downarrow_O+{}^O_\downarrow\right)_{m_{2}}\big|\cdot\cdot\cdot\big|\nonumber\\
&&-\big|\cdot\cdot\cdot\big|\left({}^\downarrow_O+{}^O_\downarrow
\right)\big|\cdot\cdot\cdot\left({}^\uparrow_O+{}^O_\uparrow\right)
\big|\cdot\cdot\cdot\big|\big]
\end{eqnarray}
and
\begin{equation}
\phi(m,m)\,=\,\big|\,\big|\cdot\cdot\cdot{}^O_{O_{{}_m}}\cdot\cdot\cdot\big|\,\big|.
\end{equation}

Define also the Fourier transforms
\begin{equation}
\phi(m,m+r)\,=\,\frac{1}{\sqrt{N}}\,\sum_{K}exp[iK(m+r/2)]\,\phi_{K}(r)
\end{equation}
for $0\leq r\leq N/2-1$ and
\begin{equation}
\phi(m,m+N/2)\,=\,\sqrt{\frac{2}{N}}\,\sum_{K}exp[iK(m+N/2)]\,\phi_{K}(N/2)
\end{equation}
The two holes are separated by a distance r. From the periodic boundary condition and 
for $r\ne N/2$, the allowed values of K are $K=(2\pi/N)\lambda$, with $\lambda=0,1,2,...,N-1$.
For r=N/2, the allowed values of K are odd multiples of $2\pi/N$. An eigenfunction in 
the momentum space is given by
\begin{equation}
\Psi_{e}^{K}\,=\,\sum^{N/2-1}_{r=0}\,a(r)\,\phi_{K}^{r}
\end{equation}
where K is an even multiple of $2\pi/N$. When K is an odd multiple of $2\pi/N$,  
the eigenfunction is $\Psi_0^K$ and the sum in Eq.(13) runs from 0 to N/2. The exact 
eigenvalue equations for both the cases are given in Ref.[13]. When K is an 
even multiple of $2\pi/N$, the amplitudes $a(r)$ have the form
\begin{equation}
a(r)\,=\,sin[q(N/2\,-\,r)]\,\,for\,1\leq\,r\leq\,N/2\,-1
\end{equation}
The energy eigenvalues are obtained by simultaneously solving the equations
\begin{equation}
\epsilon\,=\,2\,T\,cos\,q
\end{equation}
\begin{equation}
\epsilon\,+\,\frac{3J}{4}\,=\,\frac{4\,T^2}{\epsilon+\frac{3J^\prime}{4}\,-\,2\,t^\prime}\,+\,\frac{T\,sin[q(N/2\,-\,2)]}{sin[q(N/2\,-\,1)]}
\end{equation}
where $\epsilon =E-3J^\prime/2+2t^\prime$ and ,as before, energy E is measured w.r.t. that 
of the ground state in the undoped case. The energies for real values of q 
correspond to free hole states. Energies for bound and antibound states 
are obtained by making q complex. When T is +ve, making the changes $q\rightarrow\,iq$
and $q\rightarrow\pi+iq$, one gets the energies  for antibound and bound states, respectively.
When T is negative, the reverse is true. Similar results are obtained when K is an 
odd multiple of $2\pi/N$.

We now study the eigenvalue problem in the limit $N\rightarrow\infty$. 
The continuum of hole excited states, for real q, is given by Eqn.(15). For
complex q bound and antibound states are obtained. Let us now replace q by $\pi+iq$
in Eqns.(15,16). Since N is large, Eqn.(16) reduces to
\begin{equation}
\epsilon\,+\,\frac{3J}{4}\,=\,\frac{4\,T^2}{\epsilon\,+\,\frac{3J^\prime}{4}\,-\,2\,t^\prime}\,-\,T\,e^{-q}
\end{equation}
From a simultaneous solution of Eqn.(15) (with q replaced by $\pi+iq$) and Eqn.(17),
one gets the following cubic equation in $e^q$,
\begin{equation}
e^{3q}\,-\,e^{2q}\,[\frac{3J\,+\,3J^\prime}{4T}\,-\,\frac{2\,t^\prime}{T}]\,+\,e^q\,[\frac{3J}{4T^2}(\,-\,2\,t^\prime\,+\,3J^\prime/4)\,-\,3]\,-\,(\frac{3J}{4T})=0
\end{equation}
The exact, analytic solutions of a cubic equation are given in Ref.\cite{Murray}. For a 
physical solution, $e^q$ is greater than or equal to 1. There are at most 
two physical solutions of the cubic equation in (18). Once a solution for $e^q$
is obtained, the energy eigenvalue is obtained from Eqn.(15) (with q 
replaced by $\pi+iq$). For positive values of T, one gets the solution for a 
bound state of two holes and for T -ve, a solution for the antibound state is 
obtained. The other values of the excitation branches are obtained 
by symmetry. Fig.2 shows the exact energy spectrum for the bound state, 
continuum of scattering states and antibound states of two holes for $J=t=t^\prime=1$ 
and $J^\prime=2J$. Fig.3 shows the same for the parameter values $J/t=0.25$,
$t=t^\prime=1$ and $J^\prime=2J$. The bound state of holes is obtained
irrespective of the value of J/t being less than or greater than 1.
 Dagotto et al \cite{Dagotto3} were the first to show the binding
of two holes in a two-chain ladder system. Their finding was based on exact 
diagonalization of finite-sized ladder systems. Later, Troyer et al \cite{Troyer}    
also found evidence for the binding of holes in finite ladder systems. In the 
case of our model, we have shown exactly and analytically the binding of two holes
for $N\rightarrow\infty$. For finite systems also, one can solve the eigenvalye 
problem exactly. 

The two-hole ground state is the bound state of two holes with centre of mass 
momentum wave vector K=0. The exact bound state wave function is given by (13)
with K=0 and q replaced by $\pi+iq$ in (14). In the limit $N\rightarrow\infty$, one obtains
\begin{equation}
\frac{a(n)}{a(0)}\,=\,(-1)^{(n-1)}\,e^{-(n-1)q}\,\frac{a(1)}{a(0)}
\end{equation}
This result shows explicitly that the bound state wave function has an 
exponential decay as the separation between the two holes increases. With 
the knowledge of the eigenvalue $\epsilon$, the ratio $\frac{a(1)}{a(0)}$ can be 
computed from the exact eigenvalue equations derived in Ref.[13]. Fig.4 
shows a plot of $|\frac{a(r)}{a(0)}|^2$ versus r for the ground state wave function 
with parameter values $J=t=t^\prime=1.0$ and $J^\prime=2J$(dotted line), $J^\prime=10J$ (solid line).  
When $J^\prime$ is much larger than J , the holes prefer to be on the same rung to 
minimise the loss in exchange interaction enrgy. The hole delocalization 
energy along the rung is, however, lost. When $J^\prime$ and J are comparable,  
$|\frac{a(r)}{a(0)}|^2$ has maximum value when holes are separated by approximately 
one lattice constant. The exchange energy loss is less when two holes 
are on NN rungs than when they are further apart. Being on separate rungs, 
the holes gain in delocalization energy. The bound state is also more
extended. These results are in agreement with the numerical results of Troyer
et al \cite{Troyer}.

The low energy modes of a ladder system are characterised by their spin. 
Singlet and triplet excitations correspond to charge and spin modes 
respectively. The two hole ground state is in the singlet sector and, 
as already discussed, corresponds to a bound state of two holes for K=0.    
Since a hole bound state branch exists in the singlet sector, excitations 
with energy infinitesimally close to the ground state energy are possible. 
These excitations are the charge excitations since the total spin is still 
zero and the charge excitation spectrum is gapless.

There are two distinct types of spin excitations. The first is the magnon (S=1) 
excitation of the undoped ladder with energy $J^\prime$ measured with respect to the ground 
state energy. The spin triplet excitation appears on doping the ladder. 
For a pair of holes, the lowest triplet excitation energy is $-4t-2t^\prime+3J^\prime/2$ 
from Eqn.(8). The lowest triplet excitation energy depends on the values of t, 
$t^\prime$ and $J^\prime$. The spin gap energy $\Delta_{SG}$ is the difference in energies of the
lowest triplet excitation and the ground state (two hole bound state in 
the singlet sector) energy. Fig.5 shows $\Delta_{SG}$ versus J/t for $t=t^\prime=1.0$ 
and $J^\prime=2J$. Thus, the two-chain ladder model has the feature that the charge
excitation is gapless but the spin excitation has a gap. The same result holds
true for the conventional spin ladder \cite{Rice,Troyer}. In the notation
CxSy \cite{LB}(x gapless charge and y gapless spin excitations), the t-J type
ladder model exists in the C1S0(Luther-Emery) phase.


The experimental evidence of hole based superconductivity \cite{Maekawa} in a 
ladder system provides the motivation to look for superconducting pairing 
correlations in our ladder model. We have already shown the existence of the
two-hole bound state. Define the pairing operator
\begin{equation}
\Delta_{ij}\,=\,c_{i\downarrow}\,c_{j\uparrow}\,-\,c_{i\uparrow}\,c_{j\downarrow}
\end{equation}
and consider the quantity 
\begin{equation}
\tilde{\Delta}_{ij}\,=\,\langle\,2|\Delta_{ij}|0\,\rangle
\end{equation}
$|0\rangle$ and $|2\rangle$ are the ground states in the case of zero and two holes
respectively. For our ladder model, both of those ground states are exactly 
known and one can verify that ${\tilde{\Delta}}_{i\,i+\hat{x}}$ and ${\tilde{\Delta}}_{i\,i+\hat{y}}$ have opposite
signs, $\hat{x}$ and $\hat{y}$  denote unit vectors in the x(along chain) and y(along rung) 
directions. This is a signature of d-wave pairing and shows that the bound 
state of two holes has symmetry of the d-wave type. In the case of cuprate 
superconductors, there is much experimental evidence that the pairing wave
function has d-wave symmetry \cite{Scalapino}.

In the large $J^\prime$ limit, the ladder model can be mapped onto an effective boson 
model \cite{Troyer}. The physical picture is that of bound hole pairs existing 
along rungs and moving in a background of rung spin singlets. The hole 
pairs can be considered as hard core bosons. The pair hopping matrix element 
to second order in perturbation theory is   
\begin{equation}
t_b\,=\,\frac{2\,t^2}{\frac{3J^\prime}{4}\,-\,2\,t^\prime}
\end{equation}
There is also an interaction $V_b$ between two hole pairs on NN rungs. To second 
order in perturbation theory, 
\begin{equation}
V_b\,=\,\frac{4\,t^2}{\frac{3J^\prime}{4}\,-\,2\,t^\prime}
\end{equation}
Both $t_b, V_b << J^\prime$ and one can map the ladder model onto an effective hardcore
boson model on a chain with NN interaction:
\begin{equation}
H_{eff}\,=\,-\,t_{b}\,\sum_{i}(b_{i}^{\dag}b_{i+1}\,+\,H.C.)\,+\,V_b\sum_{i}n_{i}n_{i+1}
\end{equation}
${b_i}^{\dag}$ is the hard core boson creation operator, creating a hole pair on the rung 
i and $n_i$=${b_i}^{\dag}$ $b_i$ is the corresponding number operator. There is a well known mapping 
between the effective boson model and the quantum XXZ spin model in a magnetic 
field \cite{LJ}, the Hamiltonian of which is given by 
\begin{equation}
H_{xxz}\,=\,\sum_{i}[J_zS_{i}^{z}S_{i+1}^z\,+\,J_{xy}(S_{i}^xS_{i+1}^x\,+\,S_{i}^yS_{i+1}^y)] \,-h\,\sum_{i}S_i^z
\end{equation}
The operator transformations connecting $H_{eff}$ and $H_{xxz}$ are
\begin{equation}
b_j\,=\,S_{j}^{\dag}\,\,,b_{j}^{\dag}\,=\,S_{j}^{-}\,\,,n_j\,=\,1/2\,-\,S_{j}^z
\end{equation}
There is a one-to-one correspondence between the phases of the spin model and 
those of the boson model. The disordered paramagnetic phase corresponds to the 
metallic phase for charged bosons. The AFM $Ne^\prime el$-type order in the z 
direction (when $J_z\,>\,J_{xy}$) describes the ordering of bosons on the 
lattice. For charged bosons, one obtains an insulating charge-ordered phase. 
The transition from the paramagnetic to the AFM phase represents a 
metal-insulator transition. The AFM XY order $(J_{xy}>J_z)$ is characterised by
 a two-component order parameter and in the bosonic language corresponds to the 
 off-diagonal long range order of a superfluid condensate. For charged bosons, 
 this is the SC phase.

For the XXZ chain, the asymptotic forms of the correlation functions have 
been obtained by Luther and Peschel using bosonization theory \cite{Luther}.
For $|\frac{J_z}{J_{xy}}|\leq 1$, the expressions for the correlation functions in the 
limit of large x and zero magnetic field are:
\begin{equation}
<S^z(x,t)S^z>\,\sim\,cos(2k_Fx)\,x^{(-1/\theta)}
\end{equation}
\begin{equation}
<S^{\dag}(x,t)S^->\,+\,<S^-(x,t)S^{\dag}>\,\sim\,x^{-\theta}
\end{equation}
where the exponent
\begin{equation}
\theta\,=\,\frac{1}{2}\,-\,\pi^{-1}arc\,sin(J_z/J_{xy})
\end{equation}
For the equivalent bosonic model, the correlation functions
corresponding to (27) and (28) are the charge density wave(CDW) correlation
$<n_rn_0>$ and the superconducting(SC) correlation $<b_r^+b_0>$. The 
SC correlations are dominant if $\theta<1$. For our ladder model, $J_z=V_b$
and $J_{xy}=-2t_b=-V_b$, i.e., for large r both the CDW and SC
pairing correlations exist. The transformed Hamiltonian(Eqn.(25)), however, 
contains a magnetic field term. In the presence of the magnetic field 
h$(h=V_b)$, the spin chain with $|J_z/J_{xy}|=1$ is in a spin-flop phase \cite{Johnson}
which is equivalent to the SC phase in the bosonic theory. Thus for our 
ladder model, the SC pairing correlations are dominant for large $J^\prime$.

\section*{III.Conclusion}

We have considered a two-chain t-J ladder model for which several exact,
analytical results can be derived for the case of two holes. Inclusion of 
the diagonal exchange and hopping terms enables us to reduce the original 
N-body ( N-2 spins and two holes) problem to an effective two-body problem
which is easily solved. The ground state is a bound state of two holes with 
centre of mass momentum wave vector K=0 and total spin S=0. The bound state
wave function has modified d-wave symmetry. The charge
excitation is gapless whereas the spin excitation has a gap. All the results
derived by us are in agreement with the numerical results for the conventional
two-chain spin-ladder. In the strong coupling limit, our results are the only 
exact, analytical results for the lightly doped two-chain t-J ladder. For 
more than two holes, we have not been able, as yet, to calculate the ground state 
and low lying excitation spectrum exactly and analytically.

Recently, in a remarkable paper \cite{Lin}, Lin, Balents and Fisher have studied
weakly interacting electrons hopping on a two-chain ladder. Using bosonization 
and perturbative renormalisation-group(RG) analysis, they have shown that at 
half-filling the model scales onto the Gross-Neveu(GN) model. The GN model 
happens to be integrable and has SO(8) symmetry. For repulsive interactions,   
the two-chain ladder exhibits a Mott insulating phase at half-filling with 
d-wave pairing correlations. The exact energies of all the low-lying excited 
states can be calculated because of the integrability of the GN model.
Lin et al further studied the effects of doping a small density of holes into 
the d-Mott spin liquid phase at half-filling. Again, for a pair of holes, 
the ladder system exists in a SG phase with hole binding in the ground
state and gapless charge excitations.
Scalapino, Zhang and Hanke \cite{Zhang} have considered the strong coupling 
limit of a two-chain ladder model with local interactions designed to exhibit 
exact SO(5) symmetry. This model too has a SG phase with  hole pairs in
the ground state. Numerical calculations on the
t-J \cite{Troyer} and Hubbard ladders \cite{Noack} also show the existence of 
such a phase. Thus, the SG phase with bound hole
pairs appears to be a universal feature of the two-chain ladder system 
irrespective of the strength of the coupling. This phase also exhibits
superconducting pairing correlations. For ladder systems the existence of a SG
is favourable for the binding of holes. As mentioned in the Introduction, the
existence  of a `pseudo-SG' in the cuprates is conjectured to be associated
with pre-formed Cooper pairs of holes. This conjecture is supported by our
rigorous demonstration that the ground state in the SG phase consists of a
bound pair of holes.

\newpage
\section*{Figure Captions}
\begin{description}
\item[Fig. 1]The spin ladder model described by 
the $t-t^\prime-J-J^\prime$ Hamiltonian (Eqn.1).
\item[Fig. 2] Exact energy spectrum ($\epsilon$ vs K) for the bound state, continuum
 of scattering states and anti-bound states of two holes ($J=t=t^\prime=1, J^\prime=2J$).
\item[Fig. 3] Exact energy spectrum ($\epsilon$ vs K) for the bound state, continuum
of scattering states and anti-bound states of two holes ($J=0.25,J^\prime=2J,t=t^\prime$=1).
\item[Fig.4] A plot of $|\frac{a(r)}{a(0)}|^2$ vs. r for the ground state wave function
of two holes (Eqn.(13)) ($J^\prime=2J$ (dotted line), $J^\prime=10J$ (solid line)).
\item[Fig. 5] The spin gap $\Delta_{SG}$ vs. J/t ($t=t^\prime=1.0, J^\prime=2J$).
\end{description}
  
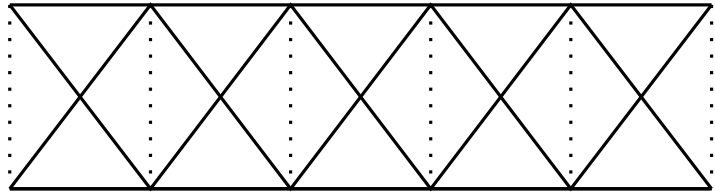
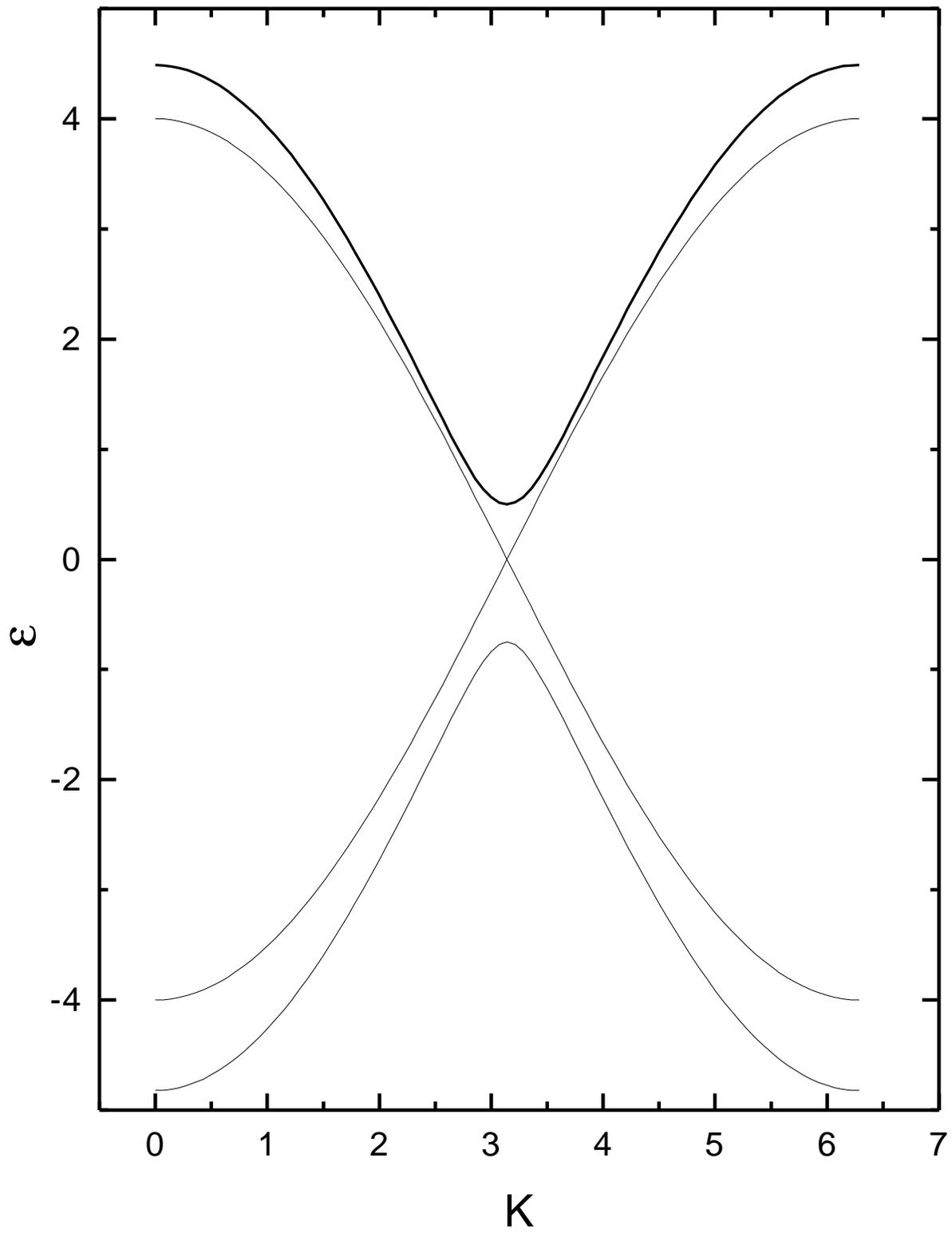
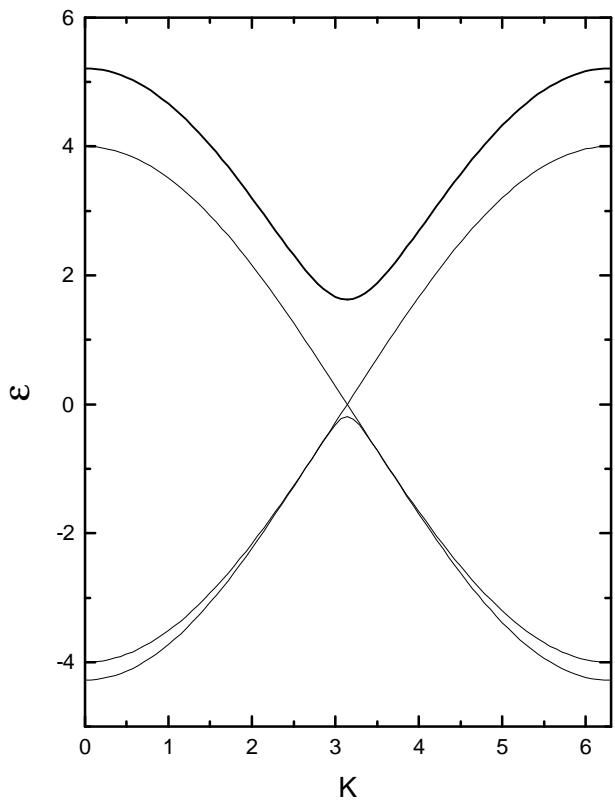
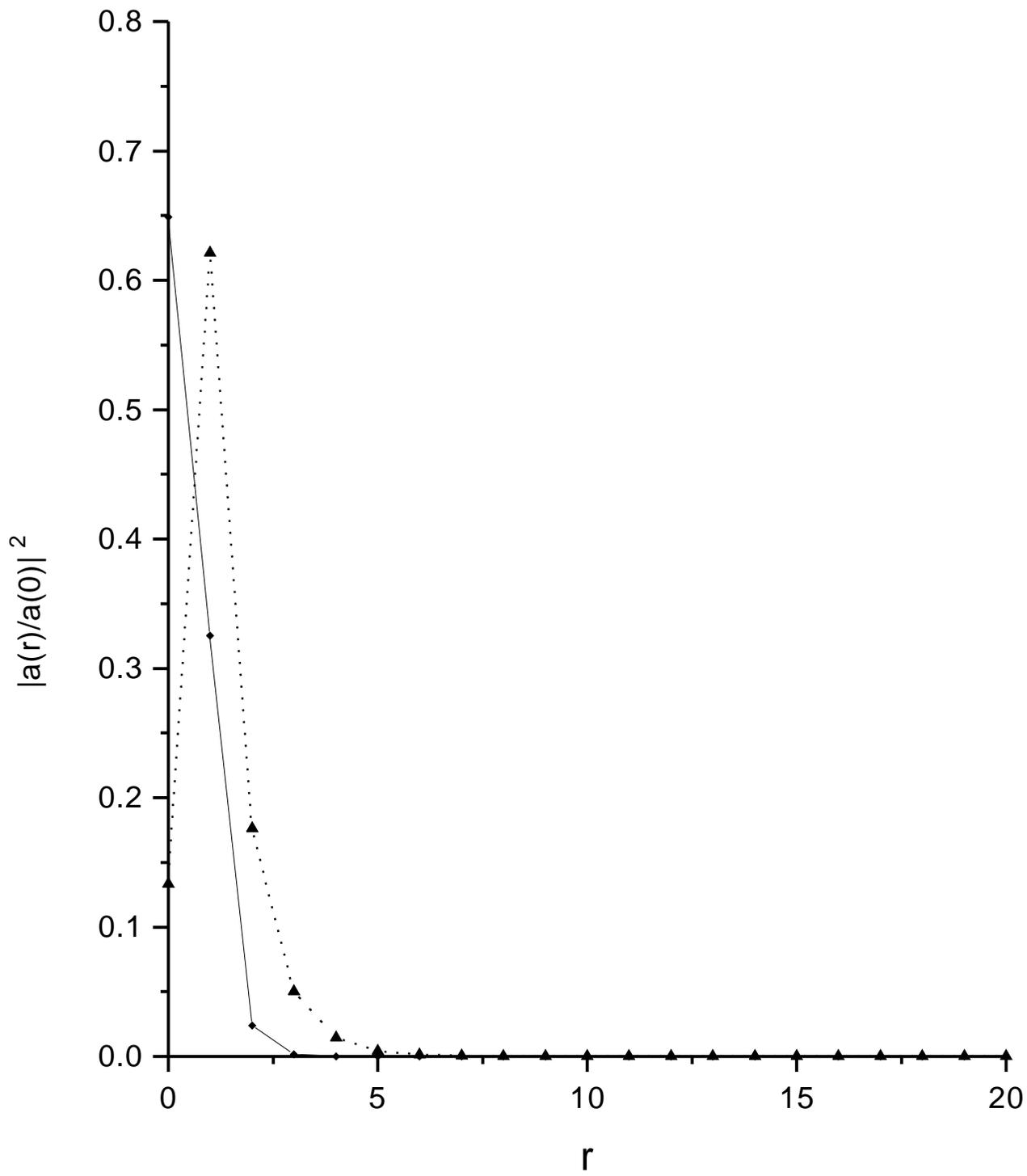
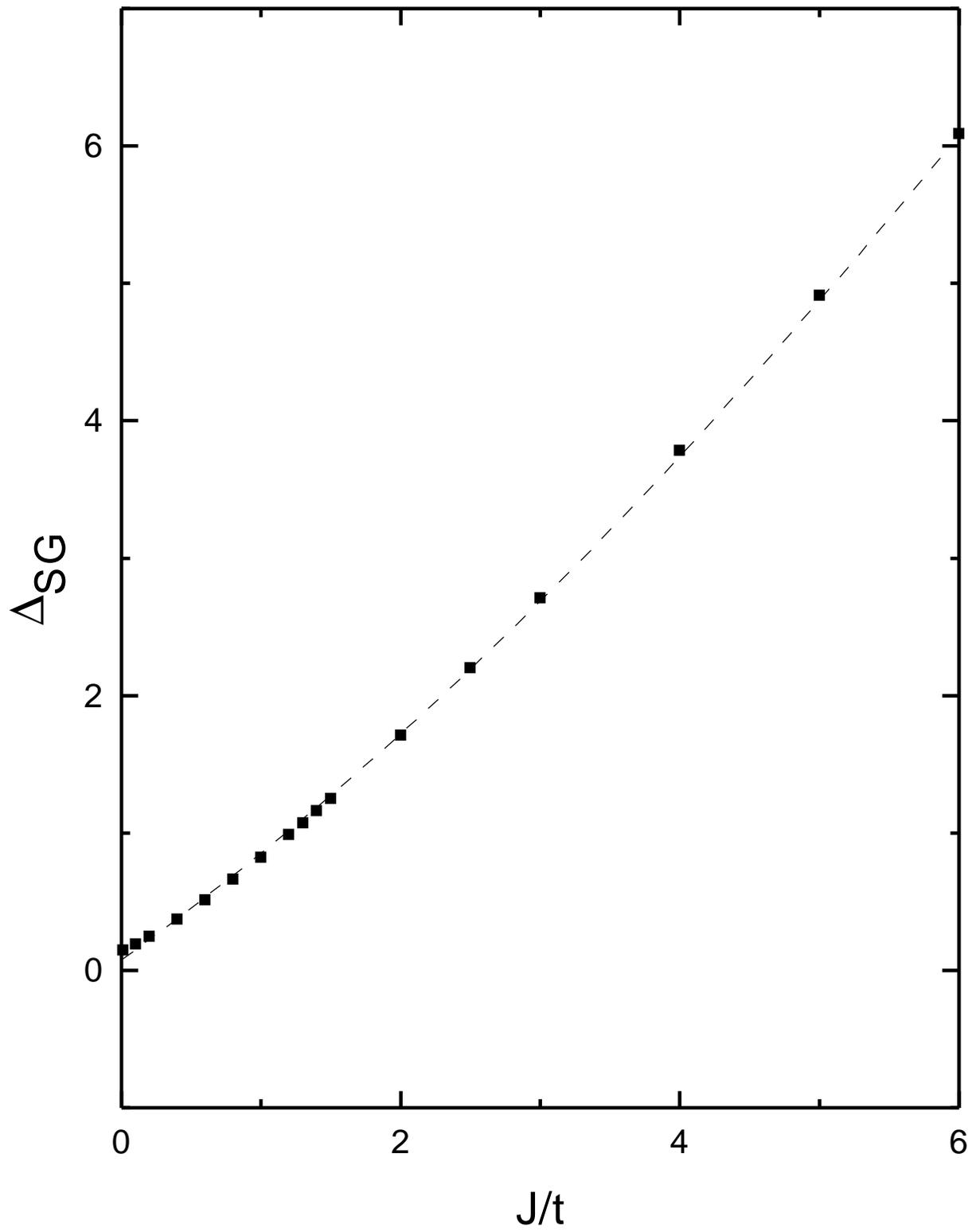
\end{document}